\documentclass[twocolumn,showpacs,preprintnumbers,amsmath,amssymb]{revtex4}
\usepackage{amssymb}
\usepackage[dvips]{graphicx}
\begin{document}

\title{ Hall-Lorenz number paradox in cuprate superconductors}
\author{A. S. Alexandrov}

\affiliation{Department of Physics, Loughborough University,
Loughborough, United Kingdom}

\begin{abstract}
Significantly different normal state Lorenz numbers have been found
in two independent direct measurements  based on the Righi-Leduc
effect, one about 6 times smaller and the other one about 2 times
larger than the Sommerfeld value in single cuprate crystals of the
same chemical composition. The controversy is resolved in the model
where charge carriers are mobile lattice bipolarons and thermally
activated nondegenerate polarons. The model numerically fits several
 longitudinal and transverse kinetic coefficients
 providing a unique explanation of a sharp maximum in
the temperature dependence of the normal state Hall number in
underdoped cuprates.
\end{abstract}


\maketitle

Particular interest in studies of high-temperature superconductors
lies in a possible violation of the Wiedemann-Franz (WF) law in
doped cuprates. A departure from the Fermi/BCS liquid picture is
seen in both the superconducting and normal state thermal
conductivities and might be related by a common
mechanism\cite{TAKE,HILL}.  Takenaka et al.\cite{TAKE}
systematically studied the oxygen-content dependence of  the
insulating state thermal conductivity enabling them to estimate the
phononic contribution, $\kappa_{ph}(T)$ for the metallic state to
some extent.   Their analysis led to the conclusion that  the
electronic term, $\kappa$ is only weakly $T$-dependent. This
approximately $T$-independent $\kappa$ in the underdoped region
therefore implies the violation of the WF law since the resistivity
is found to be a non-linear function of temperature in this regime.
A breakdown of the WF law has been seen in other cuprates such as
$Pr_{2-x}Ce_xCuO_4$  at very low temperatures\cite{HILL}. On the
other hand measurements by Proust et al.\cite{PROU} on
$Tl_{2}Ba_{2}Cu0_{6+\delta }$ have suggested that the
Wiedemann-Franz law holds perfectly well in the overdoped region.
However in any case the extraction of the electronic thermal
conductivity has proven difficult and inconclusive as $\kappa$ and
 $\kappa_{ph}$ are comparable at elevated
temperatures, or there is a thermal decoupling of phonons and
electrons at ultra-low temperatures\cite{smith}.

 This uncertainty
has been avoided in measurements of the Righi-Leduc effect. The
effect describes transverse heat flow resulting from a perpendicular
temperature gradient in an external magnetic field, which is  a
thermal analog of the Hall effect. Using the effect the
"Hall-Lorenz" electronic number, $ L_{H}=\left( e/k_{B}\right)
^{2}\kappa _{xy}/(T\sigma _{xy})$ has been directly
measured\cite{ZHANG} in $YBa_{2}Cu_{3}O_{6.95}$ and
$YBa_{2}Cu_{3}O_{6.6}$
since transverse thermal $%
\kappa _{xy}$ and electrical $\sigma _{xy}$ conductivities involve
presumably  only  electrons. The experimental $L_{H}(T)$ showed a
quasi-linear temperature dependence above the resistive $T_{c}$,
which strongly violates the WF law. Remarkably, the measured value
of $L_{H}$ just above $T_{c}$ turned out precisely the same as
predicted by the bipolaron theory\cite{NEV}, $L=0.15L_{0}$, where
$L_{0}=\pi^2/3$ is the conventional Sommerfeld value. The breakdown
of the WF law  revealed in the Righi-Leduc effect\cite{ZHANG}  has
been explained by a temperature-dependent contribution of thermally
excited single polarons to the transverse
magneto-transport\cite{leeale}.

 Surprisingly more recent measurements of the Hall-Lorenz number
in single crystals  of optimally doped $YBa_{2}Cu_{3}O_{6.95}$ and
optimally doped and underdoped $EuBa_{2}Cu_{3}O_{y}$ led to an
opposite conclusion\cite{mat}. The experimental $L_H$ for  these
samples has turned out only weakly temperature dependent and
exceeding the Sommerfeld value by more than 2 times in the whole
temperature range from $T_c$ up to the room temperature.  Following
an earlier claim\cite{li} Matusiak and Wolf\cite{mat} have argued
that a possible reason for such significant difference
 might be that Zhang et al.\cite{ZHANG} used different samples, one for $\kappa_{xy}$ and
another one  for $\sigma_{xy}$ measurements, which makes their
results for $L_H$ inconsistent.

Here I argue that there is no inconsistency in both $L_H$
determinations. One order of magnitude difference in two independent
direct measurements of the normal-state Hall-Lorenz number
 is consistently explained by the bipolaron theory\cite{alebook}. The theory explains the huge difference in the
Hall-Lorenz numbers by taking into account the difference between
the in-plane resistivity of detwinned\cite{ZHANG} and
twinned\cite{mat} single crystals. The theory fits well the observed
$L_H(T)$s  and explains a sharp Hall-number maximum\cite{mat}
observed in the normal state of underdoped cuprates.

In the presence of the electric field \textbf{E}, the temperature
gradient $\boldsymbol\nabla{T}$ and a weak magnetic field
\textbf{B}$\parallel$ \textbf{z} $\perp$ \textbf{E} and
$\boldsymbol\nabla{T}$, the electrical currents in $x,y$ directions
are  given by
\begin{eqnarray}
j_{x}\mathbf{=}a_{xx}\nabla_x(\mu-2e\phi)+a_{xy}
\nabla_y(\mu-2e\phi) \notag
\\
+b_{xx}\nabla_xT+b_{xy}\nabla_yT,
\notag
\\
j_{y}\mathbf{=}a_{yy}\nabla_y(\mu-2e\phi)+a_{yx}\nabla_x(\mu-2e\phi)
\notag
\\
+b_{yy}\nabla_yT+b_{yx}\nabla_xT,
\notag
\\
\end{eqnarray}
and the thermal currents are:
\begin{eqnarray}
w_{x}\mathbf{=}c_{xx}\nabla_x(\mu-2e\phi)+c_{xy}\nabla_y(\mu-2e\phi)
\notag
\\
+d_{xx}\nabla_xT+d_{xy}\nabla_yT
\notag
\\
w_{y}\mathbf{=}c_{yy}\nabla_y(\mu-2e\phi)+c_{yx}\nabla_x(\mu-2e\phi)
\notag \\
 +d_{yy}\nabla_yT+d_{yx}\nabla_xT.
 \notag
 \\
\end{eqnarray}
Here $\mu$ and $\phi$ are the chemical and electric potentials.

Real phonons and (bi)polarons are well decoupled in the
strong-coupling regime of the electron-phonon
interaction\cite{alebook} so the standard Boltzmann equation for the
kinetics of renormalised carriers is applied. If we make use of the
$\tau(E) -$approximation\cite{ANSE}  the kinetic coefficients of
bipolarons
 are found as\cite{leeale} ($k_B=\hbar=c=1$)
\begin{eqnarray}
a^{b}_{xx}&=&a^{b}_{yy}=\frac{2en_b}{m_b}\langle\tau_b\rangle,\notag
\\
a^{b}_{yx}&=&-a^{b}_{xy}=\frac{2eg_bBn_b}{m_b}\langle\tau_b^2\rangle,
\notag
\\
b^{b}_{xx}&=&b^{b}_{yy}=\frac{2en_b}{Tm_b}\langle(E-\mu)\tau_b\rangle,
\notag
\\
b^{b}_{yx}&=&-b^{b}_{xy}=\frac{2eg_bBn_b}{Tm_b}\langle(E-\mu)\tau_b^2\rangle,
\notag
\end{eqnarray}
and
\begin{eqnarray}
c^{b}_{xx}&=&c^{b}_{yy}=\frac{n_b}{m_b}\langle(E+2e\phi)\tau_b\rangle,\notag
\\
c^{b}_{yx}&=&c^{b}_{xy}\frac{g_bBn_b}{m_b}\langle(E+2e\phi)\tau_b^2\rangle,
\notag
\\
d^{b}_{xx}&=&d^{b}_{yy}=\frac{n_b}{Tm_b}\langle(E+2e\phi)(E-\mu)\tau_b\rangle,\notag
\\
d^{b}_{yx}&=&-d^{b}_{xy}=\frac{g_bBn_b}{Tm_b}\langle(E+2e\phi)(E-\mu)\tau_b^2\rangle,
\notag
\end{eqnarray}
where
\begin{equation}
\langle Q(E) \rangle=\frac{\int_{0}^{\infty}dE Q(E) E
D_{b}(E)\partial f_{b}/\partial E}
{\int_{0}^{\infty}dEED_b(E)\partial f_{b}/\partial E},
\end{equation}
$D_b(E)\propto E^{d/2-1}$ is the density of  states of a
$d$-dimensional bipolaron spectrum, $E=K^2/(2m_b)$, $g_b=2e/m_b$,
and $f_b(E)$ is the equilibrium distribution function. Polaronic
coefficients are obtained by replacing super/subscripts $b$ for $p$,
double elementary charge $2e$ for $e$ and $\mu$ for $\mu/2$ in all
kinetic coefficients,
 and $m_b$ for $2m_p$ in $a_{ij}$ and $c_{ij}$.
The kinetic energy of bipolarons, $E$ should be replaced by $E+T^*$,
where $E=k^2/(2m_p)$ is the polaron kinetic energy, and $T^*$ is
half of the bipolaron binding energy (i.e. the pseudogap temperature
in the theory\cite{alebook}).

The in-plane resistivity, $\rho$, the Hall number, $R_H$,  and the
Hall-Lorenz number, $L_H$ are expressed in terms of the kinetic
coefficients as $\rho^{-1}=2ea_{xx}$, $R_H=a_{yx}/2eB(a_{xx})^2$,
and
\begin{equation}
L_{H}=\frac{e\left[(d_{yx}a_{xx}-c_{yx}b_{xx})a_{xx}-c_{xx}(b_{xx}a_{yx}-b_{yx}a_{xx})\right]}
{2Ta_{yx}a_{xx}^2},
\end{equation}
 respectively, where $a,b,c,d=a^p+a^b,b^p+b^b,c^p+c^b,d^p+d^b$.

The $in$-plane resistivity, the temperature-dependent paramagnetic
susceptibility, and  the Hall ratio have  already been  described by
the bipolaron model taking into account thermally activated single
polarons\cite{alebra,jung,in,alezavdzu}. The bipolaron model has
also offered a  simple explanation of  $c$-axis transport and the
anisotropy of cuprates\cite{alekabmot,hof2,in,zve}. The crucial
point is that single polarons dominate in  $c$-axis transport at
finite
temperatures because they are much lighter than bipolarons in  $c$%
-direction. Bipolarons can propagate across the planes only via a
simultaneous two-particle tunnelling, which is much less probable
than a single polaron tunnelling. However, along the planes polarons
and inter-site bipolarons propagate with  comparable effective
masses\cite{alebook}. Hence in the mixture of nondegenerate
quasi-two-dimensional (2D) bosons and thermally excited 3D fermions,
only fermions  contribute to $c$ -axis transport, if the temperature
is not very low, which leads to the thermally activated $c$ -axis
transport and to the huge anisotropy  of cuprates\cite{alekabmot}.

We have also shown\cite{leeale} that by the necessary inclusion of
thermally activated polarons, the model, Eq.(4)  predicts a
breakdown of the WF law with the small near-linear in temperature
 Hall-Lorenz number, as observed experimentally by Zhang et
al.\cite{ZHANG} (see Fig.1). Let us now show that the bipolaron
model  describes the contrasting observations of Ref.\cite{mat} as
well, if the ratio of bipolaron and polaron mobilities,
$\alpha=2\tau_b m_p/\tau_p m_b$ becomes relatively small.

 Both polaronic and bipolaronic carriers are not degenerate above
 $T_c$, so the  classical distribution functions,
 $f_b=y\exp(-E/T)$ and $f_p=y^{1/2} \exp[-(E+T^*)/T]$ are applied with
 $y=\exp(\mu/T)$. The chemical potential is evaluated
using $2n_b+n_p=x/v_0$, where $x$ is the number of itinerant holes
in the unit cell volume $v_0$  not localised by disorder. The
bipolaron density remains large compared with the polaron density in
a wide temperature range, so that  $n_bv_0\approx x/2$ and $y
\approx \pi x/(m_ba^2T)$ for quasi-2D bipolarons. Then the atomic
density of 3D polarons  is $n_pv_0=Tm_pa^2 \exp(-T^*/T)
 (xm_p/2\pi^2m_b)^{1/2}$ ($a$ is the lattice constant). The ratio $\beta=n_p/2n_b$ remains small at any
 pseudogap temperature $T^*$ and any relevant doping
 level $x > 0.05$, $\beta \approx T exp(-T^*/T)(18 m_p/\pi^2 xm_b)^{1/2}/W \ll 1$, if the temperature $T$ is
 small compared with  the polaron bandwidth $W=6/m_pa^2$. Hence, if the
 mobility ratio $\alpha$ is of the order of unity,
 both longitudinal and transverse in-plane magnetotransport is
 dominated by bipolarons, which explains a remarkably low $L_H$ in
 high-quality detwinned crystals used in Ref.\cite{ZHANG}, Fig.1.

{\begin{figure}
\begin{center}
\includegraphics[angle=-90,width=0.50\textwidth]{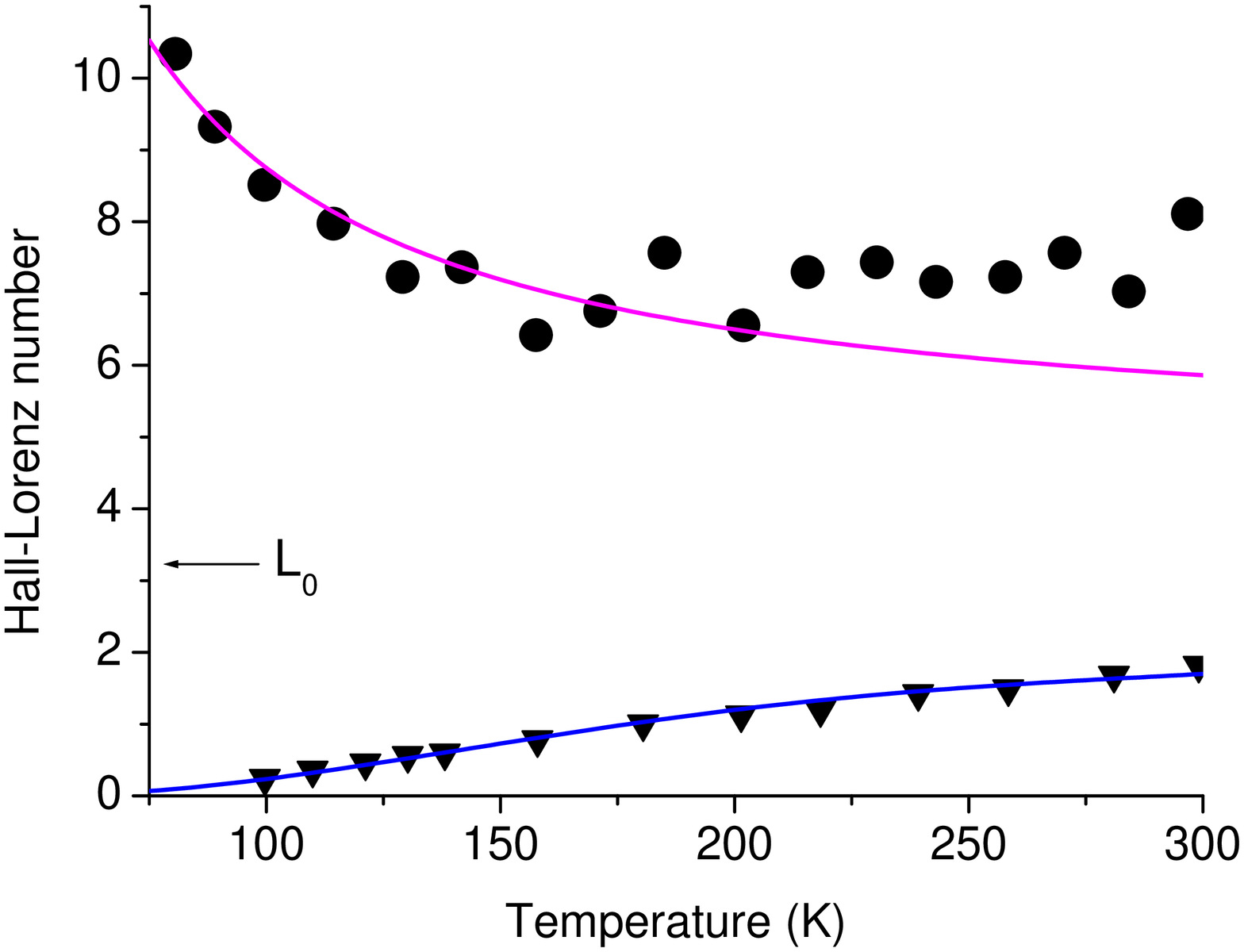}
\vskip -0.5mm \caption{The Hall-Lorenz number $L_H$ in underdoped
twinned $EuBa_{2}Cu_{3}O_{6.65}$  (circles)\cite{mat} compared with
the theory, Eq.(7) when $\alpha \ll 1$(upper line), and  the
significantly different Hall-Lorenz number in detwinned
$YBa_{2}Cu_{3}O_{6.95}$  (triangles)\cite{ZHANG} described by the
same theory\cite{leeale}  with a moderate value of
 $\alpha=0.44$ (lower line).}
\end{center}
\end{figure}

 On
 the other hand, twinned crystals used in Ref.\cite{mat} had the
 in-plane resistivity several times larger than those of Ref.\cite{ZHANG}
 presumably resulting from  twin boundaries and long
 term aging. The twin boundaries and other  defects are
 strong scatterers for slow 2D bipolarons (see below), while lighter quasi-3D polarons
are mainly scattered by real optical phonons, which are  similar in
all crystals. Hence one can expect that $\alpha$ becomes small in
twinned crystals of Ref.\cite{mat}. If the condition $\alpha^2 \ll
\beta$ is met, then only polarons contribute to the transverse
electric and thermal magnetotransport. It explains about the same
thermal Hall conductivities ($\kappa_{xy} \approx 2.5\times 10^{-3}
B$ W/Km at T=100K) dominated by polarons in both crystals of
$YBa_{2}Cu_{3}O_{6.95}$ used in Ref.\cite{ZHANG} and in
Ref.\cite{mat},   and at the same time a substantial difference of
their electrical Hall conductivities, $\sigma_{xy}$, as bipolarons
virtually do not contribute to $\sigma_{xy}$ in the twinned samples.

To arrive at simple
 analytical results and illustrate their quantitative agreement with the experiment\cite{mat} let us  assume
 that $\alpha^2 \ll\beta$, but $\alpha \gtrsim \beta$,
 and neglect an energy dependence of the transport relaxation rates of all carriers.
  In such conditions
 bipolarons  do not contribute
to transverse heat and electric flows, but determine the in-plane
conductivity.  Kinetic responses are grossly simplified as

\begin{eqnarray}
\rho={m_bv_0\over{2e^2x  \tau_b}}
\\
 R_H={v_0\beta\over{ex\alpha^2}}={e^3 n_p  \tau_p^2\over{m_p^2}}\rho^2,
\\
L_{H}=4.75+3T^*/T+(T^*/T)^2.
\end{eqnarray}

As in the case of $\alpha^2 \gtrsim \beta$, discussed in
Ref.\cite{leeale}, the recombination of a pair of polarons into
bipolaronic bound states at the cold end of the sample  results in
the breakdown of the WF law, as described by two
temperature-dependent terms in Eq.(7). The breakdown is  reminiscent
of the one  in conventional semiconductors caused by the
recombination of electron-hole pairs at the cold end\cite{ANSE}.
However, the temperature dependence and the value of $L_H(T)$ turn
out remarkably different. When $\alpha^2 \ll \beta$, The Hall-Lorenz
number is more than by an order of magnitude larger than in the
opposite regime, $\alpha^2 \gtrsim \beta$. It increases with
temperature lowering rather than decreases fitting well the
experimental observation\cite{mat} in twinned underdoped single
crystals of $EuBa_{2}Cu_{3}O_{6.65}$ with $T^*=$ 100K, Fig.1. Hence
by varying the bipolaron to polaron mobility ratio, $\alpha$, the
model accounts for qualitatively different behaviours of $L_H(T)$ in
twinned and detwinned cuprates. The energy dependence of relaxation
rates might somewhat change numerical coefficients in Eq.(7), but it
does not qualitatively change the temperature dependence and the
value of $L_H(T)$.

The  Hall-Lorenz number is the ratio of different kinetic
coefficients rather than a proper kinetic response function.
However, its significant departure from the Sommerfeld value
$L_0\approx 3.3$ clearly indicates a non-Fermi liquid behaviour
since the relaxation mechanism  virtually cancels in the ratio. The
partially gapped Fermi-liquid model  used to explain large $L_H$ in
Ref.\cite{mat} predicts a quadratic decrease of $L_H(T)$ with
 temperature lowering, rather than a steep increase as observed,
Fig.1. To account for an unexpected rise of $L_H(T)$ below
$T\approx$ 160K in underdoped samples, Matusiak et al.\cite{mat}
suggested an opening of another narrower gap. However the gapped
Fermi liquid model is clearly incompatible with the near
temperature-independent resistivity  and with the sharp maximum of
the normal state Hall ratio at  100K, as measured in Ref.\cite{mat},
Fig.2. It is also hard to accept the claim of Refs.\cite{li,mat}
that the research team of Ref.\cite{ZHANG} could so badly manipulate
their data to arrive at an erroneous $L_H$   more than one order of
magnitude smaller in identical cuprates.

On the contrary our model explains the near temperature-independent
resistivity and the unusual Hall ratio, Fig.2. If we assume that in
$EuBa_{2}Cu_{3}O_{6.65}$ slow bipolarons are mainly scattered by
neutral defects and twin boundaries, their relaxation rate depends
on the temperature  as $\tau_{b0}/\tau_b =1+ (T/T_0)^{1/2}$, where
$\tau_{b0}$ is a constant. The temperature independent contribution
comes from the scattering rate off neutral impurities with the
carrier exchange\cite{ANSE} similar to the scattering of slow
electrons by hydrogen atoms. The square-root term originates in the
scattering of slow bipolarons by point defects and twin boundaries
with a temperature independent mean-free pass. The scale $T_0$  thus
depends on the relative strength of two scattering mechanisms. The
theoretical resistivity
\begin{equation}
{\rho(T)\over{\rho_0}}=1+(T/T_0)^{1/2}
\end{equation}
fits well the experimental $\rho(T)$ in the entire normal-state
region with $\rho_0=m_bv_0/(2e^2x\tau_{b0})=1.3\times$ 10$^{-5}
\Omega$m and $T_0$=321 K, Fig.2. Lighter 3D polarons are scattered
by defects and optical phonons, so that
$\tau_{p0}/\tau_p=(T/W)^{1/2}+B \exp(-\omega/T)$ with a
temperature-independent $\tau_{p0}$. Then, using $n_p \propto T
exp(-T^*/T)$, Eq.(6) yields
\begin{equation}
 R_H(T)=\rho^2(T) {A T \exp(-T^*/T)\over{[T^{1/2}+b \exp(-\omega/T)]^2}}.
\end{equation}
This expression fits extremely well the experimental $R_H(T)$ with
 temperature independent constants  $A= e^3\tau_{p0}^2(18xm_p/\pi^2m_b)^{1/2}/(v_0m_p^2)$=275 m/$\Omega^2$C
 and $b=BW^{1/2}=$122K$^{1/2}$,  the reasonable value of the characteristic optical
phonon frequency $\omega=470$K, and the same pseudogap $T^*=100$K as
in the Hall-Lorenz number in Fig.1.  It appears almost perfect even
in the critical region very close to $T_c$, Fig.2, if one uses the
experimental $\rho(T)$ in Eq.(9). However the maximum of the Hall
ratio is a normal state feature lying well above the critical region
by about 30K, Fig.2,  as in  underdoped $YBa_{2}Cu_{3}O_{y}$,
Ref.\cite{alezavdzu}. At temperatures below $T^*$ the Hall ratio
drops as the number of thermally activated polarons decreases, and
at temperatures above $T^*$ it drops since  the polaron relaxation
time decreases.
\begin{figure}
\begin{center}
\includegraphics[angle=-90,width=0.50\textwidth]{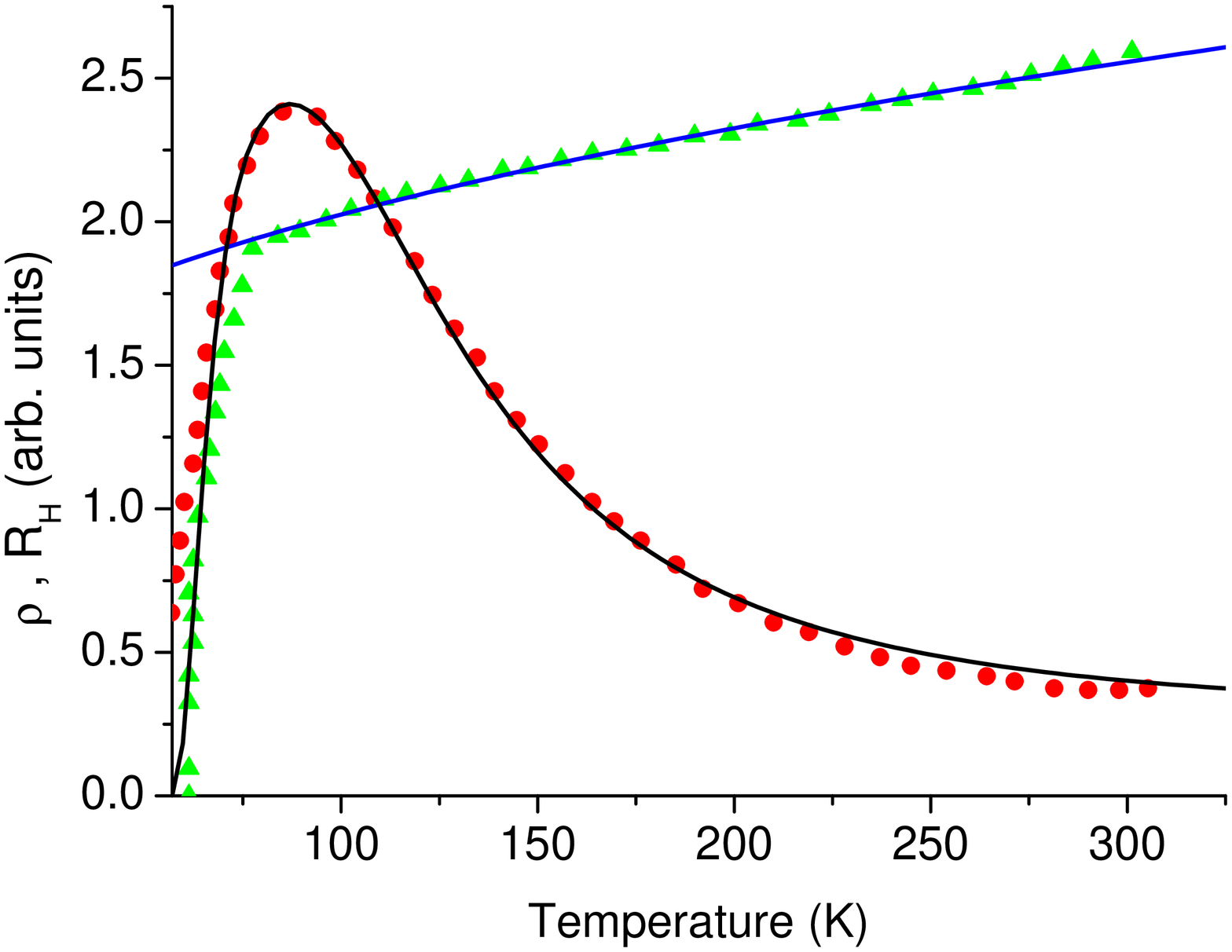}
\vskip -0.5mm \caption{The in-plane resistivity $\rho$ (triangles)
and the Hall ratio $R_H$ (circles) of underdoped  twinned
$EuBa_{2}Cu_{3}O_{6.65}$, Ref.\cite{mat},  compared with the theory
(lines).}
\end{center}
\end{figure}

To verify the self-consistency of the model let us estimate $\alpha$
and $\beta$. In the optimally doped samples one  expects $\alpha^2$
 of the order of $\beta$, so  the Hall ratio approximately
measures the itinerant carrier density, $R_H^{opt} \approx
v_0/ex_{opt}$. Then using  the experimental values\cite{mat} of
$R_H$ in optimally doped $EuBa_{2}Cu_{3}O_{7}$ and underdoped
$EuBa_{2}Cu_{3}O_{6.65}$  one estimates $\alpha^2/\beta \approx
R_H^{opt}x_{opt}/xR_H \lesssim 0.1$ in underdoped
$EuBa_{2}Cu_{3}O_{6.65}$, which justifies one of our assumptions. To
get $\alpha \gtrsim \beta$ we have to assume that $\beta \lesssim
0.1$, which is indeed the case in the whole temperature range, if
the polaron band is wide enough, $W \gtrsim 5000$K. Finally using
the values of $\rho_0$ and $A$ and taking $x=0.1$, $m_p=5m_e$,
$v_0=0.2$ nm$^3$ and $m_b = 2m_p$ the polaron and bipolaron
mean-free pass is estimated as $l_p \approx $ 4 nm and $l_b\approx
0.3(m_b/m_e)^{1/2}$nm, respectively (here $m_e$ is the free electron
mass). Their values are large compared with the lattice constant
justifying the Boltzmann approximation for all carriers.

To sum up, the bipolaron theory resolves the paradox of very
different Hall-Lorenz numbers found in two independent
measurements\cite{ZHANG,mat} in cuprate single crystals. It explains
a flat temperature dependence of the in-plane resistivity and the
sharp maximum in the normal-state Hall number of underdoped cuprates
as well.

The author acknowledges support of this work by EPSRC (UK) (grant
EP/C518365/1) and enlightening discussions with Nigel Hussey of
thermal conductivity measurements.

\end{document}